\documentclass[a4paper,10pt,twocolumn]{article}
\usepackage{graphicx,amsmath, amsfonts, amssymb, amsthm}

\title{Lightweight Task Analysis for Cache-Aware Scheduling on Heterogeneous Clusters (PDPTA'08)}
\author{Xavier Gr\'ehant\\{\small CERN openlab, Geneva and ENST, Paris}\\{\small CH-1211 Gen\`eve 23, Switzerland}\\xavier.grehant@gmail.com \and Sverre Jarp\\{\small CERN openlab, Geneva}\\{\small CH-1211 Gen\`eve 23, Switzerland}\\sverre.jarp@cern.ch}
\date{}

\begin{document}
\maketitle

\begin{abstract}
We present a novel characterization of how a program stresses cache. This characterization permits fast performance prediction in order to simulate and assist task scheduling on heterogeneous clusters. It is based on the estimation of stack distance probability distributions. The analysis requires the observation of a very small subset of memory accesses, and yields a reasonable to very accurate prediction in constant time.
\end{abstract}
\section{Introduction}
Heterogeneous resources bring in clusters the opportunity for workload placement optimizations \cite{Pieper2004,Shan2004}. Cache is a core resource. The behavior of a program relative to cache determines in great part its performance on a given server. However, cache misses are difficult to predict. In order to enhance schedulers by taking into account cache resources, programs must be analyzed quickly. The program analysis overhead must not overpass the gain in scheduling efficiency.

This work is a first step towards cache-aware scheduling in heterogeneous clusters. It consists in the design and evaluation of a new program characterization. This characterization permits fast cache misses prediction. It provide the required performance for use in schedulers. It is based on the estimation of stack distance probability distributions.

Related characterizations are presented in section \ref{related}. The scope of this work is defined in section \ref{scope}. The design of the new characterization is explained in section \ref{characterization} and evaluated in section \ref{evaluation}.
\section{Related work} \label{related}
Cycle-accurate simulators return a cache event in response to each instruction. They require a handle on the application being executed \cite{Schintke2001} or an exhaustive trace of the execution \cite{Rotithor1995,Smith1982}. Although trace compression methods exist, these simulators are slow compared to other predictors \cite{Janapsatya2007}.

How well a program behaves relative to cache has been explained in the literature with the notions of program locality \cite{Phalke1995,Milutinovic2000,Fonseca2003}. Program locality has a variety of descriptions. Reducing the description size has always been a challenge for performance prediction. Programs can be decomposed into building blocks \cite{Li1999,Wolf2000,Yang2005}. Resulting descriptions are still substantial and they do not apply to all kinds of caches.

Monte Carlo performance models represent a program as inter-dependent statistical generators of stall conditions \cite{Karkhanis2004,Srinivasan2006,Srinivasan2006a}. These models are fast. The average number of cache misses in a run is correct even for complex processors. However, the cache misses generators used in these works are still specific to a cache configuration.

Fast cross-platform cache analysis is usually done using \emph{stack distances}. A \textbf{stack distance} is the number of different memory addresses accessed between two accesses to the same address. Stack distances are suited to evaluate fully associative caches with \emph{Least Recently Used} (LRU) replacement policy and with cache lines of one element. In these cases, and in the absence of pre-fetching, cache misses occur for stack distances greater than the cache size. In addition, stack distances have shown to accurately extend to set-associative caches with various cache line sizes and replacement policies \cite{Rau1977, Smith1982, Hill1989, Grimsrud1994, Brehob1999}.

For prediction, stack distances are usually recorded in a \emph{stack distance histogram}. The precision of a histogram (i.e. the range of its bins) is usually the size of a cache line. Stack distance histograms contain the number of cache misses for every cache size. Stack distance histograms are widely used for cross-platform performance prediction \cite{Marin2004,Pieper2004,Hassan2007}. They are lighter than application traces when the cache line size is known. However, their size is still substantial and the whole trace still needs to be collected.

\section{Scope of this contribution} \label{scope}
This section explains the limitations and novelty of the characterization.
\paragraph{Limitations.}
This characterization aims to predict the number of cache misses. The cost of a cache miss and the impact of pre-fetching are not studied here, although they are important to simulate and assist cache-aware scheduling. They must be addressed separately.

\textbf{Cost of a cache miss.} In modern processor architectures the cost of a cache miss on the process execution time depends on memory latency and bandwidth, the number of hardware threads, the quality of branch prediction, other platform characteristics, and on whether it occurs during direct or speculative execution. Evaluating the cost of a cache miss is not the concern of this work, which focuses on their number.

\textbf{Pre-fetching.} Modern processors use pre-fetching, a strategy that consists of loading data to cache before it is required in the program stack. Pre-fetching takes advantage of spatial locality. Along with efficient branch prediction, pre-fetching dramatically reduces the number of cache misses. However pre-fetching is externally scheduled by processors. It does not belong to cache configuration. The evaluation of how well it filters out cache misses can be done separately, as in \cite{Srinivasan2006a,Karkhanis2004}.

\textbf{Compulsory cache misses} correspond to first-time accessed memory addresses, that is, to infinite stack distances. Compulsory instruction misses are given by the binary size and compulsory data misses are given by the data size. The characterization predicts \emph{capacity} and \emph{conflict misses} according to the standard taxonomy \cite{Hennessy2006}.
\paragraph{Novelty.}
We propose a new characterization of how a program stresses cache. This characterization outperforms current methods for description size, analysis and prediction speed. It accounts for constant prediction complexity and for the fastest analysis since only small subsets of the application trace need to be extracted. It permits cross-platform cache performance prediction with reasonable to very good accuracy.

These performances are required to provide on the fly performance prediction in order to simulate and assist task scheduling on heterogeneous clusters.
\section{The characterization} \label{characterization}
We propose a characterization based on the estimation of the stack distance probability distribution. Stack distance is seen as a random variable $X$. It is fitted to a combination of well known probability distributions. The obtained distribution has a cumulative distribution function $cdf(x) = \mathbb P(X \leq x)$. If the estimation is correct, the cache misses ratio is $\mathbb P(X > cs/ls) = 1 - cdf(cs/ls)$ where $cs$ is the cache size and $ls$ the line size. The prediction is thus of constant complexity.

In addition we propose a method to refine a simple fit. Cache misses prediction requires to fit correctly only the upper values of random variable $X$. Indeed, prediction is only useful for realistic cache sizes. If one determines that no cache is smaller than a minimal cache size $ms$, then $X$ must fit the distribution correctly for values greater than $ms/ls$ where $ls$ is the line size.

The refinement algorithm is as follows. $X$ is a random variable, in fact a list of samples. $dist$ represents the parameters of a distribution, i.e. the result of a random variable fit. Function $fit$ is a regular fit. Function $fit'$ is the refined fit.
\begin{small}
\begin{verbatim}
 function bias(X, dist, ms/ls) :
    for each s in X such that s < ms/ls
       do
          s' := randomly generated from dist
       loop until s' < ms/ls
       s := s'
    end for
    return X
 end function

 function fit'(X, ms/ls) :
    dist := fit(X)
    X' := X
    for each refinement
       X' := bias(X', dist, ms/ls)
       dist := fit(X')
    end for
    return dist
 end function
\end{verbatim}
\end{small}
At each refinement, randomly generated values based on the previous estimation replace the lower samples. Suppose that an estimation minimizes the Mean Squared Error $\epsilon$. $\epsilon_{i,j}$ is the error of estimation at $i$th refinement on data at $j$th refinement. $\epsilon^{down}$ and $\epsilon^{up}$ are the contributions of lower and upper samples to the error. $\epsilon_{n+1,n+1} \leq \epsilon_{n,n+1}$ because estimation $n+1$ minimizes the error on data $n+1$. It yields $\epsilon_{n+1,n+1}^{up} + \epsilon_{n+1,n+1}^{down} \leq \epsilon_{n,n+1}^{up} + \epsilon_{n,n+1}^{down}$. Since $\epsilon_{n,n+1}^{down} = 0$ and $\epsilon_{n,n+1}^{up} = \epsilon_{n,n}^{up}$ by construction, it yields $\epsilon_{n+1,n+1}^{up} \leq \epsilon_{n,n}^{up}$. The upper samples are better fitted after each refinement.

The remaining of this paper is an evaluation of the characterization based on the analysis of SPEC CPU2006 benchmarks.
\section{Evaluation} \label{evaluation}
We instrumented SPEC binaries with PIN to obtain instructions and data load traces \cite{Pan2005}. Stack distances are extracted using the \emph{trace profiling algorithm} \cite{Hassan2007}. We developed a few tools in Java. These tools include trace analysis, estimators based on the Method of Moments for a large spectrum of distributions (Discrete, Uniform, Gamma, Generalized Pareto (GP) and Half Normal (HN)), random number generators for each of these distributions, and estimation refinement\footnote{All tools and data developed and collected for this work are available at http://code.google.com/p/mtc-project}.

The evaluation is presented in three steps. The first step shows how well stack distances fit a probability distribution. The second step shows the effect of collecting a limited number of stack distance samples. The third step is a discussion on using the analysis to predict cache misses of other parts of the program and with other input data.
\subsection{Stack distance distribution fit} \label{accuracy}
\begin{figure}[ht]\centering
\includegraphics[scale=.19]{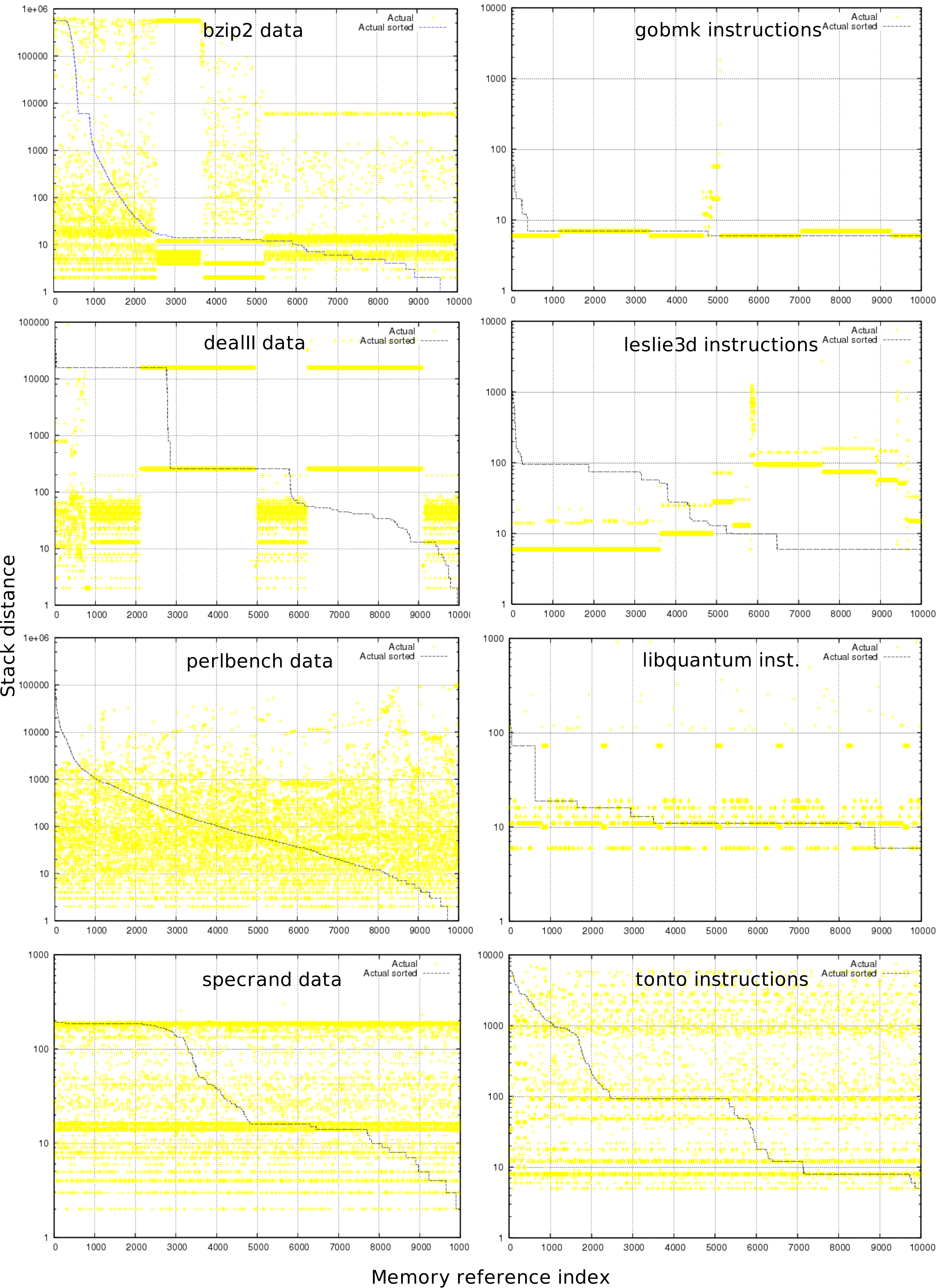}
\caption{Stack distances visualization.}\label{stacks}
\end{figure}
Figure \ref{stacks} shows stack distances on a representative cross-section of SPEC CPU2006 benchmarks. Left-hand-side figures show stack distances between data accesses, and right-hand-side figures show stack distances between instruction accesses. Light dots are stack distances in chronological order. Dashed lines are the \textbf{outlines}. They are made with the same values in descending order. The plots of figure \ref{stacks} differ from histograms. On a histogram, values are on the $x$ axis and the $y$ axis measures the number of occurrences. On figure \ref{stacks} the $x$ axis is a list of memory accesses and the $y$ axis measures corresponding stack distances.

In general, outlines are composed of curves and straight segments. An outline exclusively composed of straight segments indicates that the variable perfectly fits a discrete distribution. In this case the characterization is equivalent to estimating the histogram. It results in a compressed histogram where empty bins are removed \cite{Pieper2004}. To the contrary, a curve indicates that a histogram would require a high number of bins. When curves exist, fitting a continuous distribution dramatically reduces the characterization size, for continuous distribution is determined with typically two or three parameters. Among the 28 SPEC CPU2006 benchmarks, 11 have discrete instruction stack distances, and three (gromacs, lbm and libquantum) have discrete data stack distances. In general, a stack distance distribution is the sum of a discrete distribution and continuous distributions.

\begin{figure}[ht]\centering
\hspace{-13mm}
\includegraphics[scale=.32]{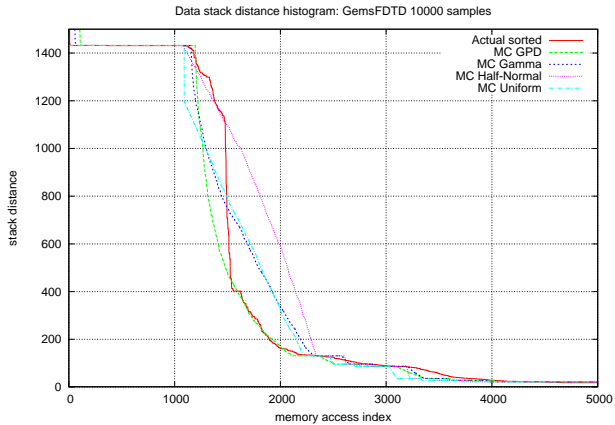}
\vspace{-13mm}
\caption{A problematic fit.}\label{fit1}
\end{figure}
Figure \ref{fit1} illustrates the analysis of GemsFDTD. The outline is shown along with Monte Carlo simulations based on different analysis. For analysis, discrete parts are filtered out and fitted separately. The remaining samples are fitted to a continuous distribution. HN fits well the upper part of the curve and GPD the lower part. However, the whole curve does not fit any single distribution alone. Gamma and Uniform average the trends. The characterization does not accurately account for all stack distances in a program whose outline has an inflexion point. 8 data traces out of the 28 benchmarks fall into this category. In these worst cases, the refined fit permits to concentrate on the higher stack distances that account for cache misses.

\begin{figure}[ht]\centering
\includegraphics[scale=.205]{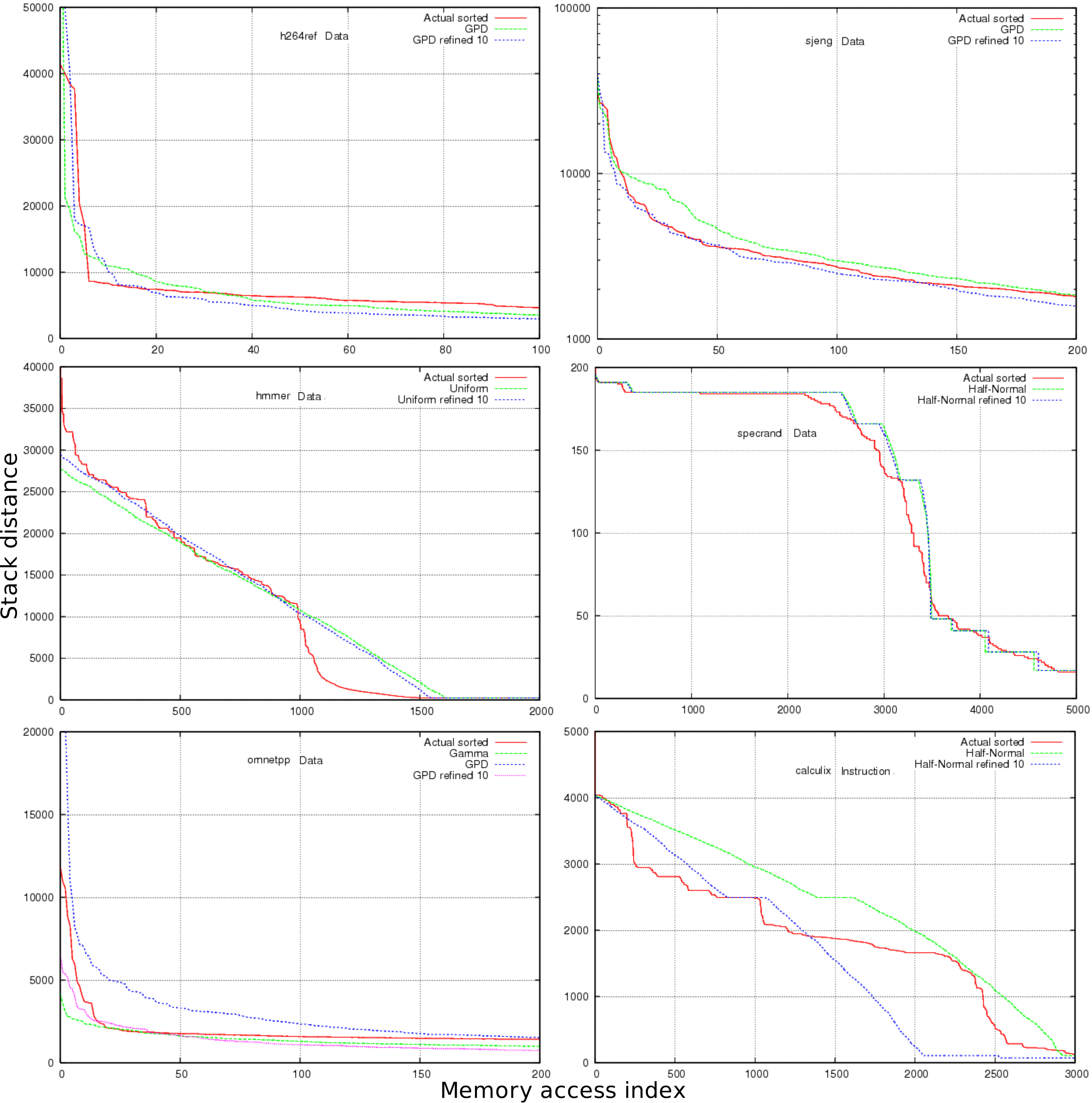}
\caption{Fit quality in various scenarios.}\label{fits}
\end{figure}
Figure \ref{fits} illustrate six representative analysis scenarios. For each benchmark the best distribution is selected, and the fit is refined. The selection can be done automatically by picking the distribution that accounts for the smallest estimation error. For the three first benchmarks, the estimation is quite accurate. Refined estimations give the best results. Indeed, outlines have long tails that bias the estimation of upper values in the absence of refinement. The precision on the fourth benchmark is impaired by the precision of the discrete fit. The two last benchmarks are the worst cases, one because of its heavy tail and the other because of its irregular outline.

In conclusion, the characterization is accurate for most SPEC CPU2006 benchmarks. Stack distances predicted in the Monte-Carlo simulations perfectly match actual values. However, there are impracticable scenarios where the precision is the order of magnitude.
\subsection{Analysis speed and prediction accuracy}
In section \ref{accuracy} we considered the ability of a probability distribution to accurately reproduce stack distances and thus predict cache misses for any kind of cache. The difference between actual stack distances and the best distribution is a first contribution to the prediction error. In this section we evaluate the number of samples required to fit such a distribution. The limited number of stack distance \textbf{samples} introduces another contribution to the prediction error. There must be just enough samples to obtain an estimation as close to the best estimation as the best estimation to the real data. In the following this number of samples is called \textbf{adequate}.

\begin{figure}[ht]\centering
\hspace{-13mm}
\includegraphics[scale=.32]{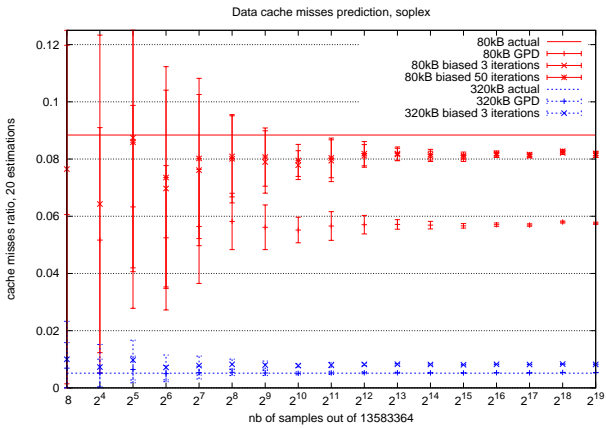}
\vspace{-13mm}
\caption{Data cache misses prediction: soplex.}\label{dsoplex80k}
\end{figure}
\begin{figure}[ht]\centering
\hspace{-13mm}
\includegraphics[scale=.32]{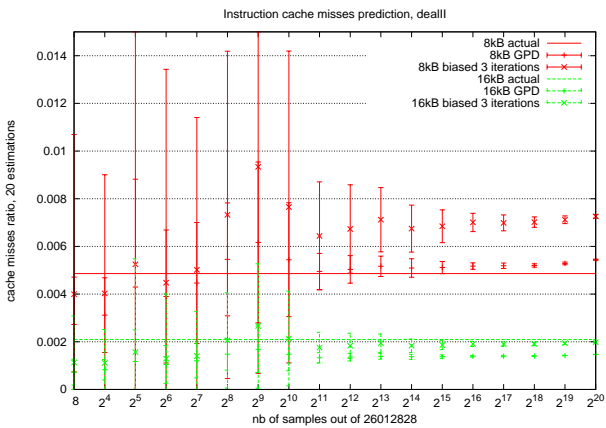}
\vspace{-13mm}
\caption{Inst. cache misses prediction: dealII.}\label{idealII8k}
\end{figure}

On figure \ref{dsoplex80k} and \ref{idealII8k}, two benchmarks are examined. Actual cache misses ratios with two different caches are compared to predictions based on different sample sets. The same characterization is used to predict cache misses for the two cache sizes.

Although refined fits are better \emph{in general}, they are not better for all cache sizes. For soplex data misses prediction with refined fits, the adequate number of samples is around $2^8$. One sample must be collected every 50,000 data accesses in memory. This number yields a prediction accuracy of 99\%. For dealII instruction misses prediction, the adequate number of samples is around $2^{11}$. One sample must be collected every 13,000 instructions, for an accuracy of 99.6\%.

In conclusion, accurate predictions are obtained with fast analysis. Less accurate predictions can be done faster.
\subsection{Prediction robustness}
In this section we briefly discuss the characterization accuracy to predict cache misses in the future and with different input data.

\begin{figure}[ht]\centering
\includegraphics[scale=.2]{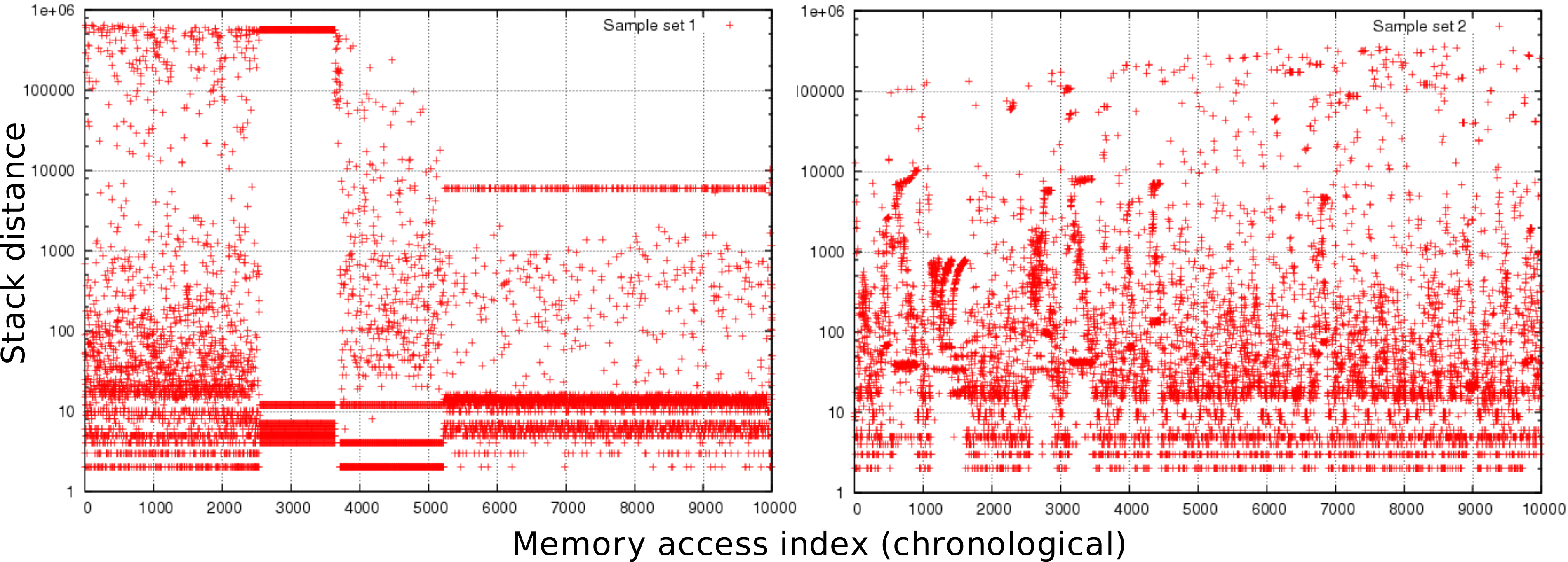}\\
\vspace{-4mm}
\hspace{-13mm}
\includegraphics[scale=.32]{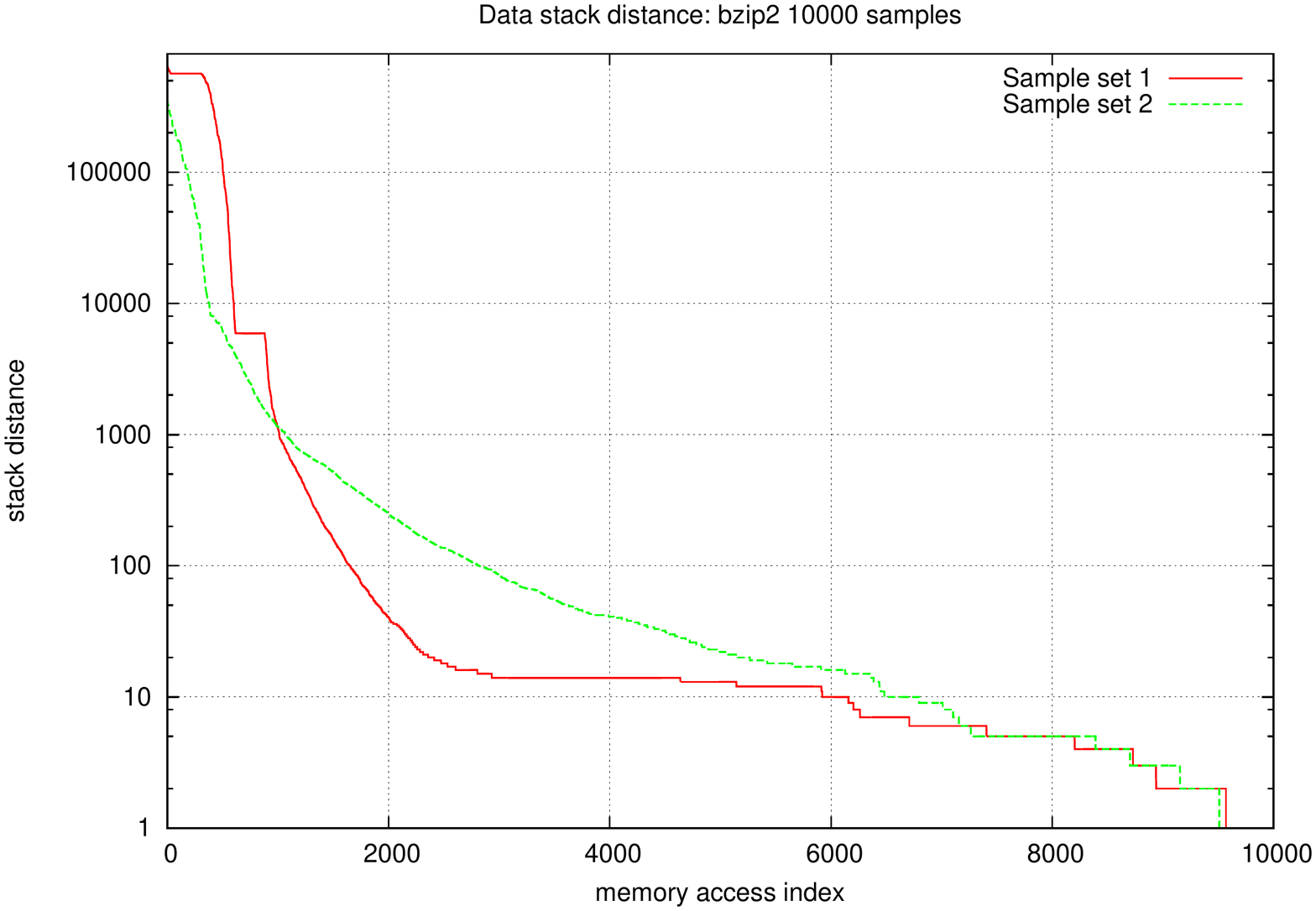}
\vspace{-13mm}
\caption{Two sample sets of bzip2.}\label{bzip2rob}
\end{figure}
Figure \ref{bzip2rob} illustrates the variation of bzip2 stack distances outline. Samples are taken from millions of consecutive memory accesses, and the two sample sets are separated by a few seconds. The samples are represented in chronological order on separate figures (top). The outlines are represented on the same picture (bottom). With bzip2 the outlines are roughly similar, but a precise prediction is not possible for future cache misses.

\begin{figure}[ht]\centering
\includegraphics[scale=.2]{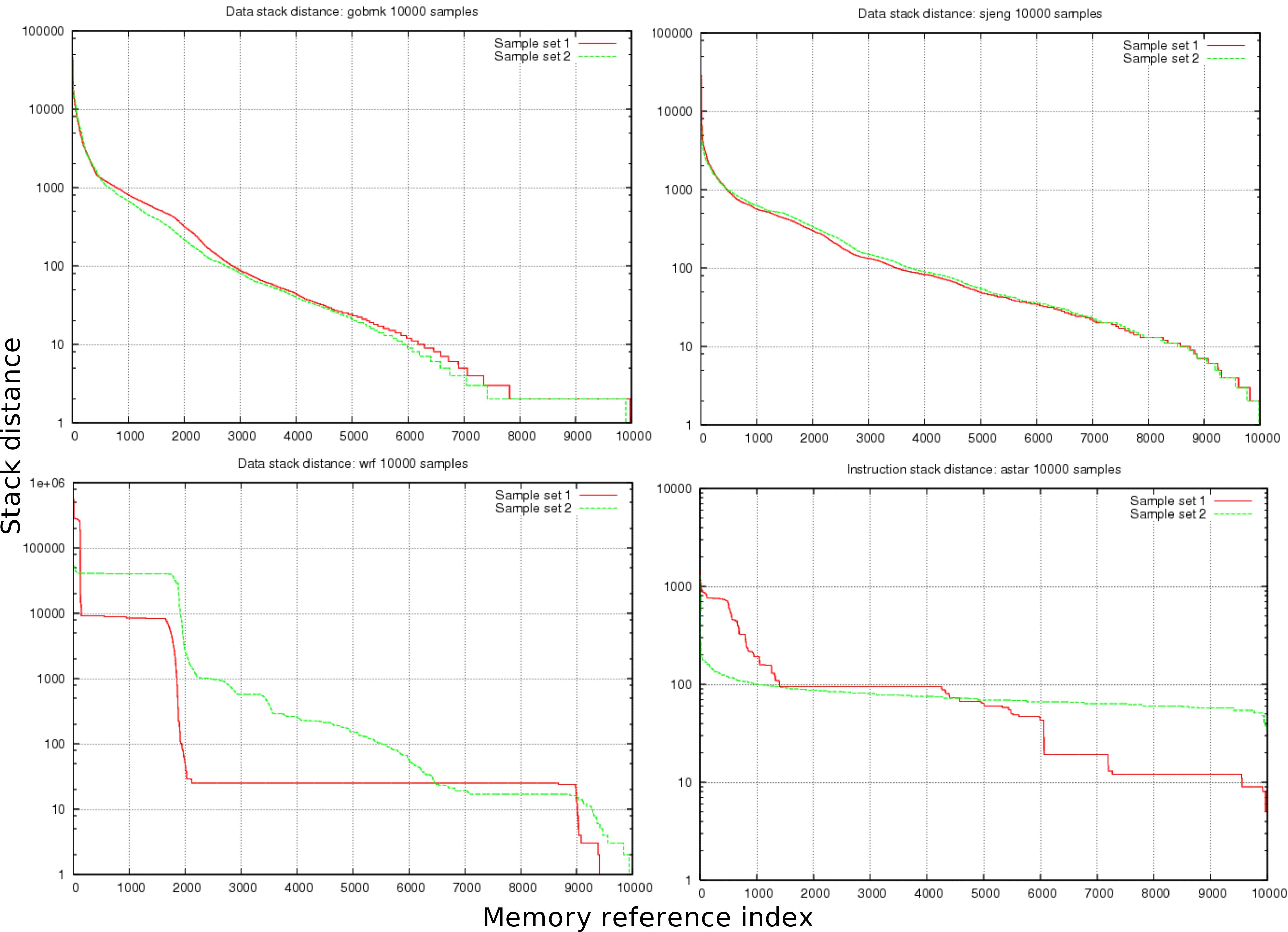}
\caption{Different inputs and observation segments.}\label{otherrobs}
\end{figure}
Figure \ref{otherrobs} shows the evolution of the outline for four benchmarks. sjeng does not change its memory access pattern in time. gobmk does not change with different input data. To the contrary, astar instruction access pattern changes in time, as well as wrf data access pattern.

In conclusion, future behaviors can be predicted only if the program is known to follow a certain regularity. For example, scientific computations often involve the repetitive execution of the same routines \cite{Yang2005}. In some cases, as with gobmk, different input data do not change memory access patterns. The analysis, unnoticeable on the first run of a routine or on the first input data, provides at worst a rough indication of the cache misses ratio, useful for cache-aware scheduling. In other cases it predicts future cache misses with very high precision.
\section{Conclusion} \label{conclusion}
We presented a novel characterization of how a program stresses cache, in terms of the stack distance fit to a probability distribution. The characterization has a very small size and provides cache misses predictions in constant time. Its evaluation distinguishes three contributions to the prediction error. One is relative to the appropriateness of a probability distribution to describe stack distances. The second is relative to the number of samples used for the fit, and the third is relative to the changes in program behavior. The worst cases yield to reasonable accuracy to simulate or assist scheduling systems. Many application behaviors are very accurately described by probability distributions and have enough regularity for the prediction to apply under different circumstances. Fitting a distribution requires the extraction of a very small subset of the trace. This makes the analysis extremely fast, which is needed to simulate and assist scheduling systems.
\bibliographystyle{abbrv}
\bibliography{xagref}
\end{document}